# Kinesthetic Weight Modulation: The Effects of Whole-Arm Tendon Vibration on the Perceived Heaviness

Keigo Ushiyama, *Member, IEEE*, and Hiroyuki Kajimoto, *Member, IEEE*

*Abstract*—**Kinesthetic illusions, which arise when muscle spindles are activated by vibration, provide a compact means of presenting kinesthetic sensations. Because muscle spindles contribute not only to sensing body movement but also to perceiving heaviness, vibration-induced illusions could potentially modulate weight perception. While prior studies have primarily focused on conveying virtual movement, the modulation of perceived heaviness has received little attention. Presenting a sense of heaviness is essential for enriching haptic interactions with virtual objects. This study investigates whether multi-point tendon vibration can increase or decrease perceived heaviness (Experiment 1) and how the magnitude of the effect can be systematically controlled (Experiment 2). The results show that tendon vibration significantly increases perceived heaviness but does not significantly decrease it, although a decreasing trend was observed. Moreover, the increase can be adjusted across at least three levels within the range from 350 g to 450 g. Finally, we discuss plausible mechanisms underlying this vibration-induced modulation of weight perception.**

## I. INTRODUCTION

THE vibration-induced kinesthetic illusion refers to an illusion of one's own body, such as position or motion [1], [2]. This illusion is typically induced by applying vibration to tendons or muscles (commonly referred to as *tendon vibration*) and has been studied to investigate movement perception and motor control [1]–[4]. It arises because vibration activates muscle spindles, which encode body position and movement (velocity and acceleration) [5]–[8]. Since this illusion can generate or enhance movement sensations without physical motion, it serves as an effective means to modulate kinesthetic perception and has been employed to create virtual reality (VR) experiences that do not require actual movement [9], [10].

Kinesthetic sensations contribute to not only movement perception but also heaviness and length of objects [11]–[13]. In other words, people can perceive these properties well by just wielding objects with their eyes closed [12]. However, the role of kinesthetic stimulation (e.g., tendon vibration) in this perception process has rarely been explored. We should therefore investigate how such stimulation influences the perception of object properties to facilitate the implementation of virtual object properties in VR environments and in teleoperation.

We previously reported that applying tendon vibration to the wrist and elbow during lifting movements could increase the sense of heaviness and resistance sensations [14], [15]. In this paper, we continue our exploration of modulating the sense of heaviness.

In this study, we reconfirmed whether tendon vibration increases perceived heaviness using a refined experimental setup and further examined its potential to decrease perceived heaviness. We also investigated whether the perceived heaviness could be systematically controlled by adjusting the vibration intensity. The hypotheses are summarized below.

H1.  The sense of heaviness can be increased by whole-arm tendon vibration, as demonstrated in our previous study [14].

H2.  The sense of heaviness can be decreased by applying tendon vibration during the movement.

H3.  The heaviness change can be adjusted by changing the strength of the illusion.

We conducted two experiments to validate these hypotheses. Experiment 1 explored whether the perceived heaviness can be decreased and whether the effect can be enhanced with three-point tendon vibration (H1 and H2). Experiment 2 examined the effect of vibration amplitudes on modulating the strength of the illusion (H3).

## II. RELATED WORK

### A. Kinesthetic illusions induced by tendon vibration

Tendon vibration is a well-established method for evoking the kinesthetic illusion. Research on how these illusions are induced began with Goodwin et al.'s study, which showed that movement illusions could be elicited by applying vibration to muscles or tendons [16]. The neural mechanisms underlying vibration-induced kinesthetic illusions have been elucidated through various approaches, including physically pulling a tendon [17], neural recordings during vibration [18]–[21], and measuring activity in the motor and somatosensory cortex [22], [23]. Because muscle spindles respond to the extension of the

This work was supported by JSPS KAKENHI Grant Number JP22KJ1370. *(Corresponding author: Keigo Ushiyama).*

Keigo Ushiyama is with the University of Tokyo, 7-3-1, Hongo, Bunkyo, Japan, Tokyo (email: keigo.ushiyama@gmail.com)

Hiroyuki Kajimoto with the University of Electro-Communications, 1-5-1, Chofugaoka, Chofu, Japan, Tokyo (email: kajimoto@kaji-lab.jp).

Approval of all ethical and experimental procedures and protocols was granted by the ethical committee of the University of Electro-Communications under Application No. H23090, and performed in line with university requirements.

muscle that contains them, vibration-induced illusions are elicited in the direction of muscle lengthening.

The strength of vibration-induced kinesthetic illusions predominantly depends on vibration parameters. Regarding vibration frequency, studies by Naito and Roll et al. found the largest illusions at 70 Hz [18], [23]. Effective frequencies for inducing the illusion have been reported to range from 50 to 110 Hz in various studies [18], [23]–[26]. In terms of vibration amplitude, Taylor et al. [1] reviewed and reported that common amplitudes ranged from 0.2 mm to 3.0 mm. While stronger vibrations generally induce more vivid illusions, it is essential to note that they can also trigger the tonic vibration reflex (TVR), leading to unintended muscle flexion [1]. Among the frequencies capable of inducing the illusion, vibration amplitudes have a greater effect on its strength than frequency [25]. In addition, the preload force of the vibration actuator influences the amplitude threshold of the illusion [27].

Meanwhile, it is generally difficult to induce a strong kinesthetic illusion. To address this, prior studies have proposed several methods to enhance the illusion for practical applications, including increasing the number of stimulated sites [26], [28], [29], combining vibration with other modalities [30], [31], and developing more effective vibration delivery techniques [32], [33]. In other words, enhancement methods are often necessary to reliably induce the illusion, especially in applications.

Most prior studies have evaluated the illusion statically (i.e., without physical movement), assessing how participants perceive position and motion illusions. However, tendon vibration remains effective even when the stimulated part is in motion. Cordo et al. [3], [34] investigated the effects of tendon vibration on the perception of velocity and dynamic position. Redon et al. [4] and Ferrari et al. [35] examined how tendon vibration during reaching tasks affects motion accuracy and performance.

### B. The functional role of muscle spindles

Muscle spindles are basically *stretch* sensors and encode the position and velocity of one's own body. Because of this role, previous studies have often evaluated the illusion with position error of the limb or velocity error of the movement.

While muscle spindles signal position and velocity to the brain during passive movement, their role can change during active movement. Dimitriou et al. measured the activity of muscle spindles (IA and II group afferent) while pressing a key [8] and grasping a block [7]. They found that the activity of the Ia group afferent was correlated with finger and wrist velocity and acceleration, and the activity of the II group afferent was correlated with velocity rather than muscle length (i.e., position). In other words, muscle spindles encoded dynamic properties (velocity and acceleration) during active motion tasks.

Muscle spindles remain active even during slow muscle *shortening*. Due to fusimotor co-activation, their firing rate increases during muscle contraction. Consequently, muscle spindles can encode not only muscle stretch but also muscle shortening through increases and decreases in firing rate [36], [37]. This suggests that signals from muscle spindles are utilized to correct motion errors and regulate force output.

### C. Force illusions by tendon vibration

Regarding the role of muscle spindles, tendon vibration has been studied in relation to position and velocity perception. However, prior work has also reported that tendon vibration can influence force perception, suggesting that muscle spindles may contribute to this modality as well.

Jones et al. [38] investigated the effect of tendon vibration on force perception by applying vibration to an elbow flexor during isometric contraction. Participants estimated the exerted force and matched it with the contralateral arm; results showed a consistent overestimation of force in the vibrated muscle. Similarly, Reschechtko et al. [39] demonstrated that vibrating finger extensors led participants to overestimate the force exerted by finger flexors.

Interestingly, both studies reported overestimation of force perception, even though one stimulated agonist muscles and the other stimulated antagonist muscles. Reschechtko et al. [39], [40] argued that afferent and efferent changes induced by tendon vibration jointly contributed to the force illusion.

### D. The sense of heaviness

The sense of heaviness is shaped through multiple modalities and related to motor efferent signals [38], [41], [42]. The size-weight illusion, in which people perceive different heaviness for objects of different sizes that have identical weights [43], is a good example of how visual information and expectations affect weight perception [44].

The sensorimotor system plays a crucial role in sensing heaviness. Previous studies have shown that perceived heaviness primarily arises from corollary discharges (a sense of effort), as evidenced by increased perceived heaviness during muscle fatigue and under local anesthesia [38]. In contrast, Davis et al. [45] demonstrated that participants unconsciously lifted larger objects with greater acceleration and peak velocity in the size–weight illusion. They argued that proprioceptive input from muscle spindles, coactivated by gamma neurons, influenced judgments of heaviness, and that a mismatch between applied force (efferent signals) and sensory feedback (afferent signals) gave rise to the illusion. This interpretation is known as the sensorimotor mismatch theory (SMT) [46].

As direct evidence that muscle spindles contribute to the weight perception, Luu et al. [47] and Brooks et al. [48] showed that muscle spindles in contracting muscles contribute to heaviness perception by desensitizing them via partial anesthesia or long-term muscle vibration. As a further exploration of muscle spindles' function, Monjo et al. [49], [50] indicated that the sense of effort can be perceived through both efferent and afferent signals, which vary with tasks. In summary, both efferent and afferent sensory signals are significant contributors to the perception of heaviness [42] regarding the sensorimotor system.

### E. Rendering heaviness in applications

Since heaviness is perceived through multiple modalities, stimulation to each modality (vision [51], skin [41], Golgi tendon organs [52], [53], and muscles [54]) has been used to simulate heaviness. Visual stimulation (i.e., pseudo-haptics) has been extensively investigated because of its compactness and the sensory dominance of vision. However, participants still perceive heaviness through proprioception, and the effect of pseudo-haptics is limited; it can break down if the discrepancy between modalities becomes too large. To address this limitation, Hirao et al. proposed combining proprioceptive stimulation (tendon vibration) with a pseudo-haptic approach for rendering heaviness [51].

Although muscle spindles play a crucial role in heaviness perception, previous studies have primarily focused on elucidating their neural mechanisms. Hirao et al. [51] examined the combined effects of proprioceptive (tendon vibration) and visual stimulation, reporting that vibration acted as perceptual noise that enhanced visual cues. In contrast, our previous study showed that tendon vibration could increase perceived heaviness without visual stimulation [14]. These findings suggest that the combination approach can work to some extent; however, it could be improved by refining the method for inducing the sense of heaviness through tendon vibration. Therefore, this study explores how tendon vibration can modulate the sense of heaviness.

### III. General Methods

This section describes the common parts of the two experiments.

### A. Stimulation methods

Because consistently applying tendon vibration to wrist and elbow flexors increased the sense of heaviness more than switching the stimulation side with the lifting direction [14], we developed Anterior and Posterior vibrations, which apply tendon vibration to the anterior or posterior side consistently during lifting. *Anterior vibration* represents a condition where vibration is applied to the tendons of arm flexors (wrist flexors, biceps brachii, and clavicular part of the deltoid) to induce arm anteversion illusion. Conversely, *Posterior vibration* corresponds to a condition where vibration is applied to the tendons of arm extensors (wrist extensors, triceps brachii, and the spinal part of the deltoid) that induce arm retroversion illusion.

We illustrated our concept of stimulation in Fig. 1. Here, our focus is primarily on velocity perception, with emphasis on the role of muscle spindles during active movement [7], [8].

First, given that lifting a heavy object involves significant resistance and effort, and participants moved their arm at constant velocity, we aimed to induce a resistance sensation during the movement to enhance the sense of heaviness. Increasing resistance against motion makes it hard to maintain constant velocity. Consequently, participants need to exert more effort to achieve the target movement, which can enhance the sense of heaviness.

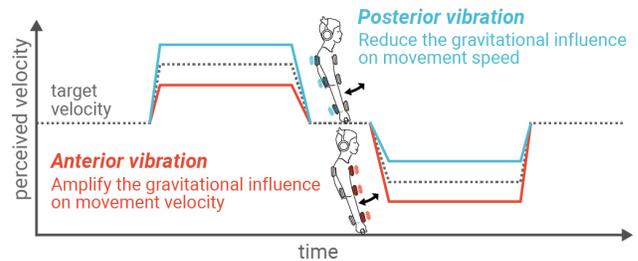

Fig. 1. Illustration depicting velocity perception changes under the two vibration conditions. Anterior vibration was designed to induce resistance during lifting and enhance perceived heaviness, whereas Posterior vibration aimed to produce the opposite effect.

Second, in the experiments, we asked participants to move their arms like lifting a dumbbell to feel heaviness. In this case, the flexor muscles on the ventral side of the arm are engaged to lift an object. By applying tendon vibration to these flexor muscles, an arm extension illusion can inherently be induced [28]. We therefore hypothesized that tendon vibration on the flexors would induce a resistance sensation during lifting (i.e., participants would perceive that the object is not moving as planned), leading to an increase in perceived heaviness.

Third, we can also induce an illusion in the opposite direction by stimulating the extensor muscles. We therefore hypothesized that an inverse effect (i.e., a decrease in perceived heaviness) could be achieved. To maintain consistency in the stimulation effect, we chose to keep applying vibration on the same side during lifting.

### B. Apparatus

Six vibration actuators (Acouve Lab, VP210) were positioned on each joint of the anterior and posterior sides, as illustrated in Fig. 2. The purpose of delivering vibration to multiple joints was to enhance the illusion [1], [28] and ensure participants experienced an illusory effect consistently. A pressure sensor (Interlink Electronics, FSR 402) was placed on the tip of each actuator to monitor the preloading force.

The vibration actuators were housed in a case (Fig. 2) and secured to the body using elastic bands (DAISO, wrist supporter, elbow supporter, and shoulder supporter). Although the actuators were placed exclusively on the right arm, elastic bands were also applied to the left elbow and wrist to provide pressure and tactile feedback similar to the right arm.

Six-point vibratory stimulation was achieved using an audio interface (Roland, OCTA-CAPTURE) coupled with three stereo audio amplifiers (FX-AUDIO-, FX202A/FX-36A PRO). Vibration signals were generated using audio-programming software (Cycling '74, Max 8).

An accelerometer (Adafruit, LIS331) was housed within each actuator case to calibrate vibration amplitudes. Additionally, six-axis sensors (Akizuki Denshi, BMX055) were attached to the participants' right and left upper arms, around the elbows, to measure the angular velocity of movement. Data from these sensors was recorded using microcontrollers (Espressif, ESP32 DevKit-C). The sampling rates were set to 1 kHz for the accelerometers and 360 Hz for the six-axis sensors.

Fig. 3 illustrates pulley equipment constructed from wooden planks, designed to allow participants to lift and compare various weights while maintaining consistent tactile feedback, as the object's size can influence perceived heaviness [11]. The grip part featured a 3D-printed hollow cylinder (30-mm diameter, 150-mm length) connected to a weight via a thread (RUNCL PowerBraid, 90lb, 0.5 mm) passing through brass pulleys.

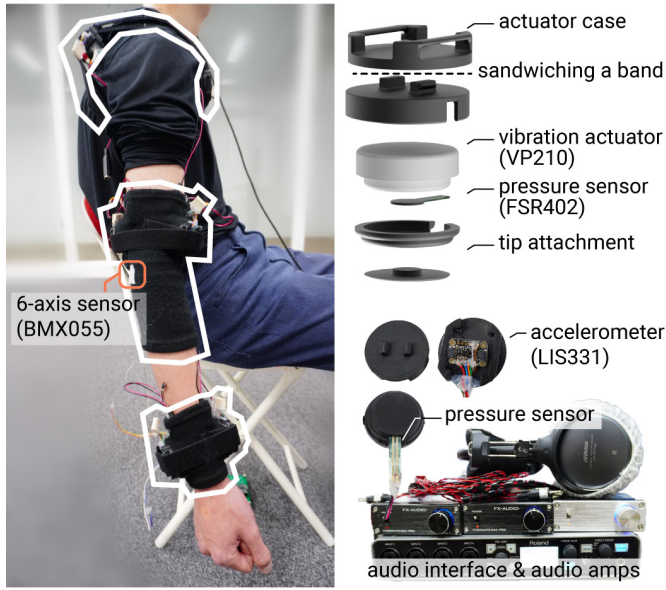

Fig. 2. Study setup for vibratory stimulation and movement sensing. The vibration actuators were housed in a 3d-printed case and fixed to elastic bands. Participants' movements were monitored using six-axis sensors attached to both upper limbs.

*C. Task*

The main task for participants was to lift weight samples using both their left and right arms with their eyes closed and determine which one felt heavier. We employed a two-arm lifting task to facilitate comparisons. Tendon vibration was consistently applied to the right arm. Due to hardware limitations (a maximum of 8 channels), we chose not to switch the stimulation side.

Lifting movement was controlled because rapid movements (approximately 70 deg/s) could potentially terminate the effects of tendon vibration [3]. Therefore, the target range of movement velocity was set between 10 deg/s and 15 deg/s. The movement profiles used in the experiments are illustrated in Fig. 4. Following auditory cues, participants performed a two-second shoulder flexion to lift the weight and a two-second shoulder extension to lower it for each comparison. Tendon vibration began with the first audio cue and ended 1 second after the movement, so participants could only sense the weights during the vibration. To maximize the effect of tendon vibration on each joint, we instructed participants to keep their elbows and wrists straight while lifting the weight with shoulder movements.

## IV. EXPERIMENT 1

The objective of this experiment is to examine whether the sense of heaviness can be increased or decreased using a different experimental setup from that used in prior work (H1 and H2).

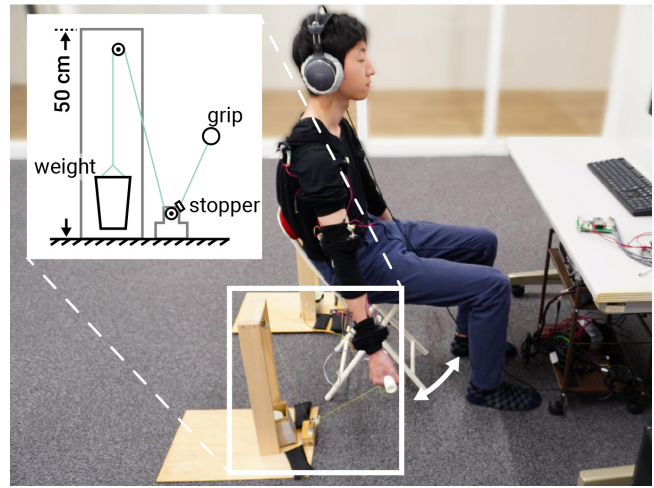

Fig. 3. Pulley equipment used in the experiments for changing the weights while keeping the gripper volume same.

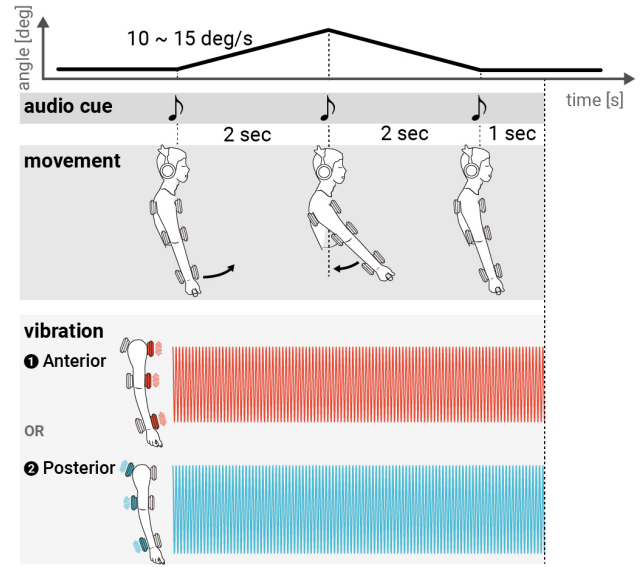

Fig. 4. Timing of audio cues and vibration relative to the lifting motion profile. *Anterior vibration* was applied to the arm flexors, and *Posterior vibration* to the extensors.

*A. Participants*

We recruited 13 participants from our university (nine males and four females, M = 23.2 years old, SD = 4.6). Eleven participants were right-handed, and two were left-handed.

*B. Stimuli*

We applied 70 Hz vibration at 70 m/s² (zero-to-peak) amplitude at each stimulation site. This frequency was chosen based on prior research indicating that the largest illusion was induced at this frequency [1]. The amplitude condition was set to the maximum amplitude that each actuator could be driven at without unintentional noise.

In addition to the two vibration conditions (Anterior and Posterior), a control condition was included in which no vibration was applied, serving as a baseline for comparing the sense of heaviness in our setup.

## C. Design

The heaviness comparison design followed the method of constant stimuli. We selected 300 g as a reference weight and prepared nine levels for comparison, ranging from 100 g to 500 g in 50-g increments (i.e., 100, 150, 200, 250, 300, 350, 400, 450, and 500 g).

The reference weight was increased by 100 g compared to our previous experiment [14]. This adjustment was made for two reasons. First, to facilitate weight comparison, as some participants reported difficulty perceiving changes in heaviness; additionally, proprioception plays a greater role in heaviness perception for heavier objects [55]. Second, the increase created a margin below the reference weight to evaluate the reduction effect of tendon vibration. At 300 g, tactile and proprioceptive inputs contribute equally to the sense of heaviness [55], so using a heavier reference allowed us to explore proprioceptive modulation more effectively. We assumed that this change in the reference weight did not have a large impact on the overall results.

Since vibratory stimulation was consistently applied to the right arm, the point of subjective equality (PSE) was determined by varying the weight on the left arm while keeping the weight on the right arm fixed at the reference weight (300 g). This comparison task was repeated ten times for each weight sample under each stimulation condition, yielding 270 trials per participant. The trial order was pseudo-randomized.

The experiment was divided into two sessions to reduce participants' fatigue. Eight participants completed both sessions within one day, separated by more than 30-minute intervals, while 5 completed the sessions in two days. While this difference of intervals causes inclusion of the muscle fatigue effect, muscle fatigue can rapidly recover up to about 90% within 30 minutes [56]. Because the total exercise time is 9 minutes per session and tendon vibration does not affect recovery time [57], we assumed that 30 minutes would be enough to relieve fatigue. Note that this interval was extended depending on participants' fatigue.

## D. Procedure

During the installation of vibration actuators on participants' bodies, each actuator was secured with additional rubber bands while the pressure forces were monitored by the pressure sensors. The preload forces of each actuator were adjusted to fall within the range of 1 N to 3 N. This adjustment aimed to minimize the illusion threshold [27] while ensuring participant comfort.

We calibrated each vibration amplitude at 70 m/s² and assessed whether participants could feel the intended kinesthetic illusion. Participants were asked, "Can you feel any motion in your right arm?" based on prior work [25], [58]. If participants had difficulty perceiving the illusion, the experimenter also asked, "Can you feel a flexion (or extension) of your right arm?"

Before the comparison task, participants practiced the movement at a constant velocity (10 deg/s to 15 deg/s) until they became accustomed to it without visual feedback. During the practice session, participants held 300-gram (baseline) weights in both hands. To assist participants in timing their movements and maintaining the correct velocity, we displayed angular velocities of both arms on a front screen (only during the practice session), and sound cues were delivered through headphones at the beginning, reversal, and end of each movement (see Fig. 4). White noise was also played to mask the sound of vibration during the movements. Note that tendon vibration was not applied during this practice session.

Due to the difficulty in precisely controlling the velocity of active movement, participants were instructed to maintain the target velocity to the best of their ability. This practice lift was repeated 10 to 30 times. After the practice session, they were told to lift weights as practiced and to focus on the weight rather than precisely executing the instructed motion.

## E. Results

We confirmed that all the participants reported the kinesthetic illusion in the intended direction at the calibration phase.

One of the 3,510 (270 × 13) trials was excluded from the analysis due to an experimental error. The PSE for each condition was determined by fitting a normal cumulative distribution to the proportion of participants indicating "left weight is heavier." The results of curve fitting are depicted in Fig. 5. The average PSEs across participants were 304 g (SE = 10.1 g) for the *No vibration* condition, 427 g (SE = 19.5 g) for the *Anterior vibration*, and 270 g (SE = 20.6 g) for the *Posterior vibration* (Fig. 6). The changes from the control condition in the two vibration conditions were a 42% increase for *Anterior vibration* and an 11% decrease for *Posterior vibration*.

We also calculated the just noticeable differences (JNDs) for each condition (the difference between the 75% point and the PSE) to examine whether tendon vibration influenced perceptual resolution. The average JNDs were 20.1 g (SD = 6.79 g) in *No vibration*, 27.4 g (SD = 15.2 g) in *Anterior vibration*, and 24.9 g (SD = 15.6 g) in *Posterior vibration*. Overall, no substantial degradation of perceptual ability was observed under tendon vibration.

One-way repeated-measures ANOVA (RM-ANOVA) on the average PSEs across the conditions revealed a significant main effect of vibration condition (F(1.36, 16.3) = 25.2, $p <$ 0.001, partial $\eta^2$ = 0.678). Since the assumption of sphericity was violated (Mauchly's $W$ = 0.531, $p$ = 0.0308), the degree of freedom was corrected by Greenhouse-Geisser's epsilon ($\varepsilon$ = 0.681). A t-test corrected with the Bonferroni method found significant differences between *No vibration* and *Anterior vibration* ($t$ = -7.39, $p <$ 0.001, Cohen's $d$ = 2.05) and *Posterior vibration* and *Anterior vibration* ($t$ = 5.272, $p <$ 0.001, $d$ = 1.46). On the other hand, a significant difference was not found between the *No vibration* and *Posterior vibration* conditions ($t$ = 1.70, $p$ = 0.347, $d$ = 0.470). Although there was no statistical significance, six of 13 participants reported that the object felt lighter during *Posterior vibration*.

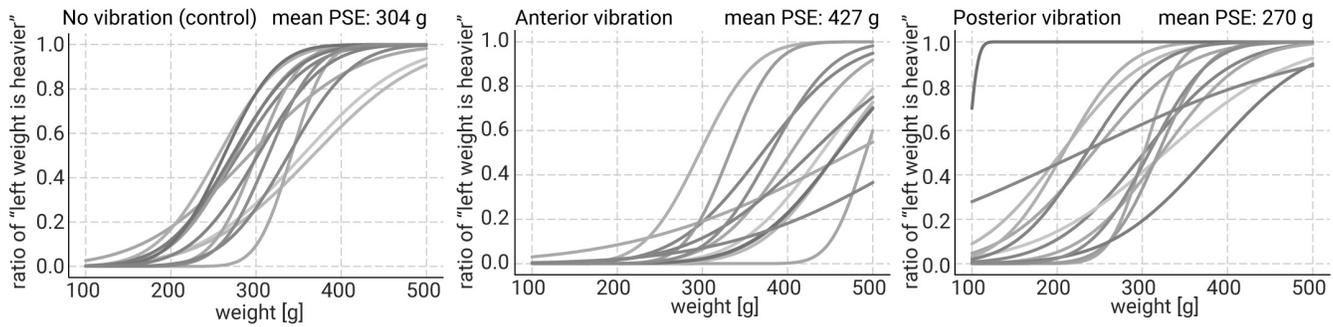

Fig. 5. Psychometric curves obtained from curve fitting for each participant. The vertical axis represents the proportion of trials in which participants judged the left weight as heavier than the right (reference) weight. The weight corresponding to a 50% response rate represents the point of subjective equality (PSE).

## F. Discussion

### 1) Soundness of the experiment

Since the control condition result was almost identical to the reference weight, the perceived heaviness was not biased toward one side. While there might have been a difference in perception capability between the right and left arms for each participant, this did not significantly affect the results. However, our results cannot exclude the effect of hand dominance and the perception difference between the right and left arms. Further studies should be conducted to better understand the mechanism of our approach by improving the study setup (e.g., automatically changing the weight samples and adjusting the vibration amplitude).

Moreover, Fig. 7 shows the results of left-handed participants (N=2) and right-handed participants (N=11) separately. Basically, no large differences are observed between *No vibration* and *Anterior vibration*. Although a difference can be observed with *Posterior vibration*, it should not be due to the difference in hand dominance because such a difference should appear with *Anterior vibration* if the hand dominance significantly affects the result.

If strong fatigue had been induced by vibration or the experimental procedure, it could have altered weight perception even in *No vibration*, as previous studies reported that fatigue can change the sense of heaviness [38], [42]. However, no such change was observed, indicating that the differences in perceived weight were due to tendon vibration.

If a TVR had occurred in the agonist muscle, it should have reduced perceived heaviness by lowering the sense of effort [42], [59]. In that case, opposite effects would have been expected in each vibration condition. Therefore, the observed changes are unlikely to result from fatigue or TVR effects.

### 2) The effect of reference weight

We increased the reference weight from our previous experiment [14] to 300 g to better elicit the effects of proprioceptive stimulation (i.e., tendon vibration) on the sense of heaviness. However, this change in reference weight may have influenced the results. In the present study, we did not systematically examine how the reference weight affects heaviness perception. To fully leverage this approach in practical applications, it is important to investigate the role of the reference weight further. For example, if the reference is varied from 100 g to 500 g, will the perceived heaviness change in a similar manner to that observed in the current experiment?

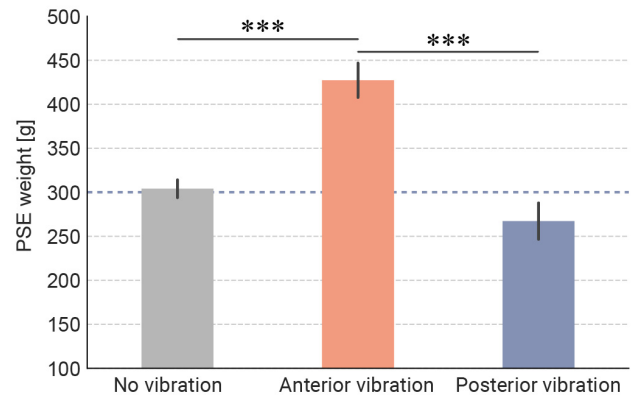

Fig. 6. Mean weights at the PSE for each vibration condition. The bold dotted line represents the reference weight (300 g). The error bars represent standard errors. ***: $p < 0.001$

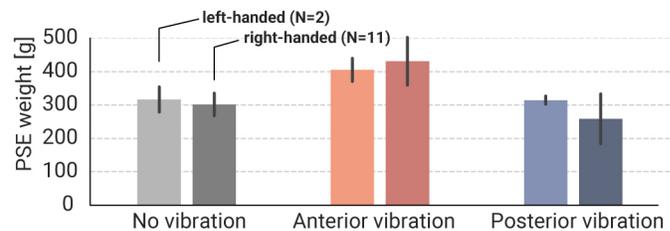

Fig. 7. The results of Experiment 1 separated by hand dominance. The error bar represents standard deviation.

### 3) Comparison with the prior study

Consistent with our previous study [14], tendon vibration increased the perceived heaviness. In our previous study, a 25% increase from the reference (200 g) was reported when wrist and elbow vibration were applied during lifting, whereas this experiment resulted in a 42% increase. While it is hard to compare the results directly, this 17% difference may be attributed to the increased number of stimulation points and the alteration in movement style (shoulder movement), as the perceived heaviness can vary based on lifting technique [60].

### 4) Limited effect of Posterior vibration on reducing heaviness

Although the average PSE in the *Posterior vibration* condition was lower than in the control (*No vibration*) condition, the difference was not statistically significant. Based on the statistical analysis, however, *Posterior vibration* had a medium effect (d = 0.47), suggesting a possible influence. Subjective reports also support this trend: five participants noted a lighter sensation, with some describing the arm as being "pushed forward" or "moving faster." These impressions imply that heaviness could decrease when vibration assists the lifting motion.

Contrary to our initial assumption (Fig. 1), *Posterior vibration* did not consistently reduce effort throughout the movement. Four participants reported that the effect reversed between the ascending and descending phases: vibration during descent produced a resistance sensation that made the object feel heavier. In addition, three participants reported little or no noticeable effect.

To summarize Experiment 1, H1 was supported and H2 was not: perceived heaviness significantly increased but did not decrease.

## V. Experiment 2

This experiment aims to investigate whether the change in perceived heaviness can be adjusted by varying the strength of the illusion (H3). Given the stable increase observed with the *Anterior vibration* condition in Experiment 1, we focused on it for this experiment.

### A. Participants

We recruited 15 participants from our university (eleven males and four females, M=24.2 years old, SD=4.9). Twelve participants were right-handed, and three were left-handed. Five out of 15 participants also experienced Experiment 1.

### B. Stimuli

In this study, the amplitude was adjusted to change the illusion strength. The vibration amplitudes were 10, 25, 40, 55, and 70 m/s². The minimum amplitude of 10 m/s² was chosen because this level rarely induces the kinesthetic illusion.

### C. Design

We measured PSE weights for each vibration amplitude using the up-and-down method with PEST [61]. PEST is a rule that controls step sizes in staircase series. In PEST, the step size is halved when the series reverses, and the series ends when the step size reaches a predefined value. The final value obtained after the last step size adjustment was used for analysis (i.e., one result per series). A notable feature of PEST is that the step size is doubled if the same response is given more than three times consecutively. This feature allows the series to track changes in the induced sensation in the participants during testing.

In this experiment, one ascending and one descending series were run at each amplitude condition (10, 25, 40, 55, 70 m/s²), starting from 200 g and 600 g, based on the range observed in Experiment 1. The initial step size was 80 g, and each series ended when the step size was reduced to 10 g (i.e., when the response reversed after a 20 g weight change). The result for each vibration condition was calculated as the average of the two series. If participants consistently responded "heavier" or "lighter" beyond the limits three times in a row, that series was exceptionally terminated, indicating that the induced sensation had saturated within the measurement range.

We chose to investigate PSE weights across multiple vibration amplitudes rather than focusing solely on the just-noticeable difference (JND), as the optimal reference amplitude and the potential magnitude of the heaviness illusion were initially unclear. Therefore, we first examined changes in perceived heaviness over a broad range of vibration amplitudes.

The order of vibration amplitudes was pseudo-randomized for each participant. Within each vibration condition, the two series were run simultaneously to prevent participants from predicting the following answer. The experimenter randomly selected a series for each trial. A one-minute interval was taken between conditions. Overall, the experiment was completed within approximately one hour.

### D. Procedure

The procedure is similar to Experiment 1. The calibration section was modified, and the illusion threshold was measured.

To control vibration amplitude in 1 m/s² increments, we created a mapping table that relates software volumes to vibration amplitudes from 1 m/s² to 70 m/s². This table was generated by gradually increasing the vibration amplitude while recording data from the accelerometers.

Next, we measured the amplitude threshold required to induce the kinesthetic illusion. Participants were asked to report whether they felt "their right arm moving backward or being moved by someone" after a four-second *Anterior vibration*. Two series—an ascending series starting from 10 m/s² and a descending series starting from 70 m/s²—were conducted simultaneously. Each series continued until the participants reversed their answers three times. The initial step size was 10 m/s², which was reduced to 5 m/s² after the first reversal. The threshold was calculated by averaging all reversal points.

At the end of the experiment, participants completed a seven-point Likert-scale questionnaire (1: strongly disagree, 4: neutral, 7: strongly agree) to assess their subjective experiences. The items are listed in Table I. The Q1 and Q2 were designed to determine where sensations induced by tendon vibration were attributed. The Q3 and Q4 aimed to evaluate whether participants were aware of the effect of kinesthetic stimulation.

Moreover, participants were asked to indicate which body parts they focused on when comparing the weights during the experiment, by encircling body parts of the body illustration shown in Fig. 9 (B). We collected the focused body parts to observe whether there is any correlation between the change of the heaviness perception and the focused body parts.

TABLE I
ITEMS IN THE QUESTIONNAIRE

| Description |
| --- |
| (Q1) I felt the hand-held object became heavier when the vibration was applied. |
| (Q2) I felt the arm became heavier when the vibration was applied. |
| (Q3) I felt resistance to the movement when the vibration was applied. |
| (Q4) I felt the resistance made the hand-held object heavier. |

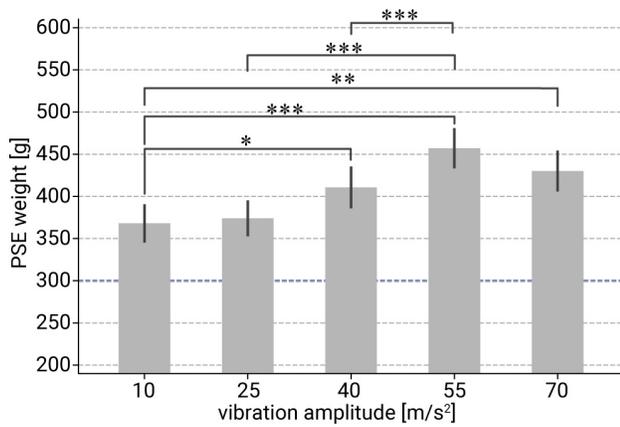

Fig. 8. Mean PSEs at five vibration amplitudes. The error bars represent standard errors. *: $p < 0.05$, **: $p < 0.01$, ***: $p < 0.001$

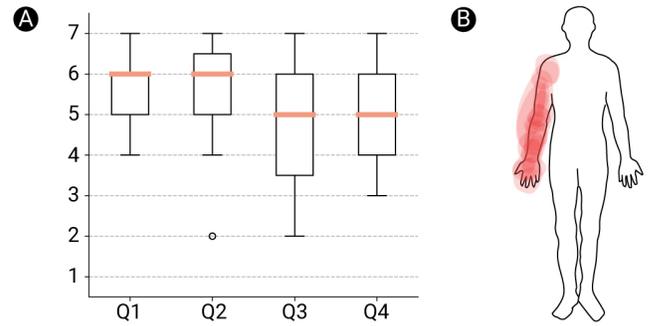

Fig. 9. (A) Box plots of questionnaire results. (B) Overlaid drawings of body parts participants focused on; only the right side is shown.

### E. Results

All participants successfully perceived the illusory shoulder extension. The average illusion threshold was 19.6 m/s² (SD = 9.15 m/s², ranging from 4 m/s² to 32 m/s²).

The results of the PSE are depicted in Fig. 8. The mean PSEs were 368 g (SE = 30.0 g), 374 g (SE = 20.4 g), 411 g (SE = 23.9 g), 457 g (SE = 23.3 g), and 430 g (SE = 22.4 g), from 10 m/s² to 70 m/s². One-way RM-ANOVA indicated a significant main effect of vibration amplitude ($F(4, 56)$ = 14.107, $p < 0.01$, partial $\eta^2 = 0.502$), with no violation of the assumption of sphericity. The paired t-test with Bonferroni correction revealed statistically significant differences among 10 m/s², 40 m/s², and 55 m/s². Also, the effect sizes are calculated using Cohen's d: the results of t-test are (1) $t$ = -3.65, $p < 0.05$, $d = 0.942$ (10 m/s² vs. 40 m/s²), (2) $t$ = -6.89, $p < 0.001$, $d = 1.78$ (10 m/s² vs. 55 m/s²), and (3) $t$ = -4.94, $p < 0.01$, $d = 1.28$ (40 m/s² vs. 55 m/s²).

The results of the questionnaire are illustrated in Fig. 9. Fig. 9 (A) shows a box plot of the four items, and Fig. 9 (B) shows the body areas where the participants paid attention when comparing the heaviness.

Although we conducted a multiple regression analysis to examine the relationship between perceived heaviness and the body parts participants focused on, no significant correlations were found. Accordingly, the analysis results are not reported here. For each participant, the drawn illustrations were converted into binary data indicating whether each of six body parts (shoulder, upper arm, elbow, forearm, wrist, and hand) was marked as a focus area, serving as the dependent variables, while the independent variable was the PSE weight. The PSE weight at 55 m/s² was chosen as the representative value because it had the highest average in Fig. 8.

### F. Discussion

#### 1) Steps for increasing the perceived heaviness

Based on the statistical analysis, the heaviness can be increased by at least three levels (10, 45, and 55 m/s²) within the range of 350 g to 450 g. Interestingly, the increase in perceived heaviness decreased at 70 m/s², contrary to our expectations. This could be due to saturation of the increase effect, TVR, or a masking effect from strong vibration.

The experiment revealed that the sense of heaviness did not increase linearly with vibration amplitude. Based on this finding, it may be possible to modulate perceived heaviness across more than three levels by identifying just-noticeable differences (JNDs) within the amplitude range of 10 to 55 m/s².

Since the average amplitude threshold of the illusion was about 20 m/s² in Experiment 2, most participants should not have perceived the illusion at the 10 m/s² condition. However, the perceived heaviness increased by about 70 g from the reference. A one-sample t-test found the average perceived heaviness at the 10 m/s² condition to be significantly higher than 300 g (t = 3.22, p < 0.01, Cohen's $d = 0.831$). While this error might be due to the study design, there may have been an offset increase caused by skin vibration [62].

If the vibrotactile effect is significant, the perceived heaviness at 25 m/s² should have increased more than at the 10 m/s² condition. However, the results at 10 m/s² and 25 m/s² almost did not differ ($t$ = -0.526, $p = 1.00$, Cohen's $d = 0.136$ in multiple comparisons). With vibration amplitudes above the illusion threshold (i.e., 20 m/s² and higher), the results began to increase linearly. Consequently, kinesthetic modulation by tendon vibration influenced the perceived heaviness, while vibrotactile stimulation also affected the perception.

The results of Experiment 2 support H3, as participants reliably perceived at least three levels of weight change within the 350–450 g range.

#### 2) Subjective evaluation

As a result of the questionnaire, participants subjectively felt an increase in the heaviness of both the object and their arms (Q1 and Q2). While there were individual differences in the interpretation of heaviness, participants generally felt that both their arm and the object became heavier.

Moreover, as shown in Fig. 9(B), a clear relationship between body parts and changes in heaviness was not found. Nine participants reported focusing on their arm (forearm, elbow, upper arm, or shoulder) rather than their hand. This fact implies that arm proprioception is strongly linked to the perceived heaviness of the hand-held object.

For the rest of the items (Q3 and Q4), the participants felt resistance during lifting by vibration and associated resistance with heaviness, while the score was lower than in Q1 and Q2. Because there is a positive correlation between Q3 and Q4 (Pearson's $r = 0.67$, $p < 0.01$), the resistance sensation by tendon vibration can be easily associated with the increase in the perceived heaviness.

## VI. GENERAL DISCUSSION

The main question is how tendon vibration alters the sense of heaviness during lifting. Weight perception arises from both efferent signals (muscle activity) and afferent signals (motion acceleration and velocity) [45], [46], [63]–[65]. To examine this, we analyzed velocity errors between the right (stimulated) and left (control) arms in Experiment 1, since participants were instructed to lift at a constant velocity; deviations from the control condition could be attributed to tendon vibration. The average error was calculated for each condition and participant using equation (1).

$$\overline{\omega}_{error}(t) = \frac{1}{N}\sum_{i=1}^{N}\left\{\omega_{right}^{(i)}(t) - \omega_{left}^{(i)}(t)\right\} \qquad (1)$$

where $\omega$ denotes angular velocity, $i$ indexes all trials in each condition (N=90), and $\omega_{right/left}^{(i)}(t)$ represents the right- or left-arm angular velocity at time $t$ for the $i$-th trial. To compute the difference, the measured angular velocities were resampled to 1800 points (5 s at 360 Hz, which is the sampling rate of the six-axis sensors used) after smoothing with a 20-sample moving average.

Fig. 10 shows representative velocity error traces from three participants. Participant #1 reported increased heaviness under *Anterior vibration* but little change under *Posterior vibration*, whereas Participants #2 and #3 experienced both increases and decreases in each condition. Because early lifting is known to be critical for heaviness perception [45], [66], we focused on the first second of movement. Fig. 11 illustrates maximum and minimum velocity errors relative to the control condition. For each participant, max/min values were calculated under each condition, and the averages across participants were then computed for each condition. Maximum and minimum values were used rather than peaks because no clear peaks were observed in some participants.

One-way RM-ANOVA revealed significant effects for both maximum and minimum values within the first second (maximum: $F(1.19, 14.3) = 15.8$, $p < 0.001$, partial $\eta^2 = 0.57$; minimum: $F(1.23, 14.8) = 24.0$, $p < 0.001$, partial $\eta^2 = 0.67$), with post-hoc tests confirming differences between conditions. If velocity errors—representing modulation of movement sensations—had caused changes in perceived heaviness (as assumed in Fig. 1), the two measures would be expected to correlate. However, no significant correlation was found between the PSE weights and velocity errors across conditions (maximum: *No vibration – Anterior*, $r = -0.13$, $p = 0.67$; minimum: *No vibration – Posterior*, $r = 0.33$, $p = 0.26$). If velocity errors were caused by TVR, the direction of the effects should have been opposite. This suggests that tendon vibration influences both velocity perception and heaviness perception.

Although *Posterior vibration* did not significantly reduce perceived heaviness, participants consistently reported changes in motion, suggesting that tendon vibration affected movement perception without robustly altering heaviness. Interestingly, one participant reported an increase in heaviness with *Posterior vibration*, and three others (including Participant #1 in Fig. 10) noted that heaviness differed between ascending and descending, with the object feeling heavier during descent. Because the stimulation was applied independent of motion, it may have acted as a perturbation or resistance, leading to a sensation of increased heaviness during certain movement phases.

The absence of a correlation between velocity errors and changes in perceived heaviness may be due to the experimental design or analysis method. Nevertheless, this implies that another mechanism may be involved. An alternative explanation is that tendon vibration directly modulated force or effort perception, rather than indirectly through movement modulation. Since sensory input from muscle spindles during contraction can contribute to heaviness perception [42], and Golgi tendon organs can also be stimulated by tendon vibration during contraction [21], such stimulation may have enhanced these signals and thereby altered weight perception.

## VII. LIMITATIONS AND FUTURE WORK

### A. Vibrotactile effect

Based on the results and discussion in Experiments 1 and 2, we believe that the kinesthetic effect of tendon vibration could stably increase and partially decrease the perceived heaviness. However, we cannot disregard the vibrotactile effect on the perceived heaviness. To further validate the mechanism of our approach, the frequency effect should be investigated using a frequency that hardly induces the kinesthetic illusion (e.g., over 200 Hz).

Kim et al. found that sinusoidal vibration on the fingers can induce a weight illusion, making people feel as if the grasped object is heavier [62], [67]. They demonstrated that an increase in grip force due to vibration correlated with the perceived heaviness [62]. Since their vibrotactile weight illusion was stronger at lower frequencies, our approach may relate to their findings.

### B. Experimental Design

To clarify the underlying mechanism of weight modulation, the heaviness changes should be compared with the intensity of motion illusions (i.e., displacement or velocity illusions). Tendon vibration has been used to induce such illusions [1], [3], as muscle spindles are primarily responsible for sensing these motion-related properties [6]. If a discrepancy is observed between the magnitudes of the motion and weight illusion, it may suggest that tendon vibration directly modulates the perception of heaviness.

### C. Stimulation methods

The proposed stimulation methods cannot be considered optimal for modulating perceived heaviness due to two main limitations. First, the effect of tendon vibration was inconsistent and appeared to depend on movement direction. Although we assumed our approach could increase or decrease the effort required to lift weights (Fig. 1), the effect varied during the ascending and descending phases, as reported by four participants. To clarify this issue, future studies should examine the effects of tendon vibration applied to agonist and antagonist muscles across different movement directions, including horizontal and vertical motion, to account for gravity effects.

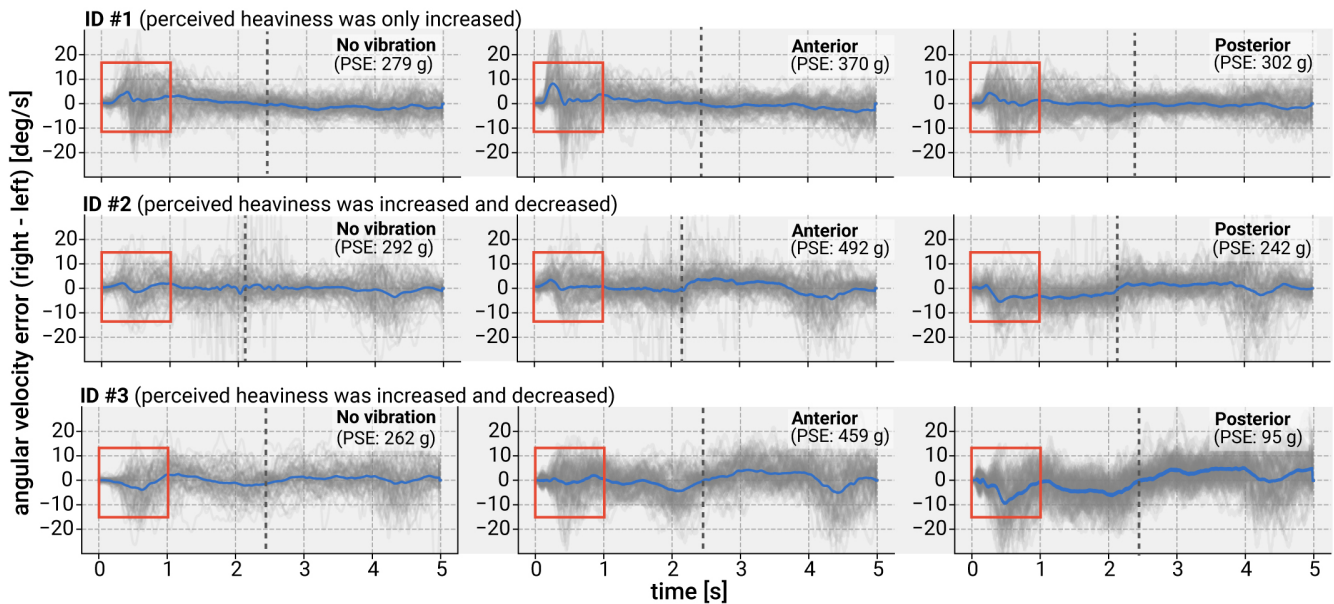

Fig. 10. Representative examples of angular velocity differences between the right and left arms. Bold lines indicate average errors, and gray lines in background show errors in individual trials. Vertical dotted lines mark the reversal from ascending to descending. Before the line, positive values indicate the right arm moved faster; after it, they indicate slower movement than the left arm.

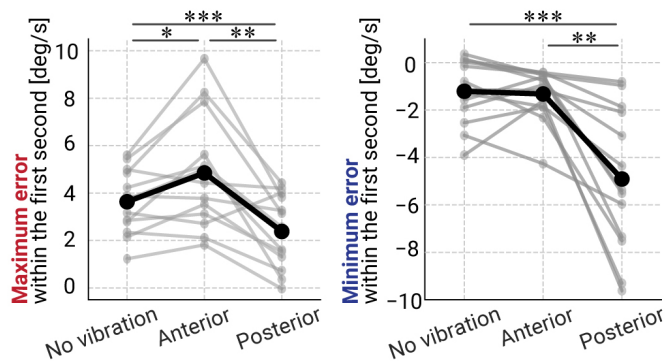

Fig. 11. Velocity errors under vibration conditions. Gray lines show individual participants; the black bold line shows the condition average. Left: maximum errors; Right: minimum errors. *: p < 0.05, **: p < 0.01, ***: p < 0.001.

Second, both movement velocity and reference weight were fixed in this study. Because the effect of tendon vibration is influenced by velocity [3] and the sensitivity of muscle spindles varies with muscle contraction due to fusimotor co-activation [1], [20], it is important to investigate whether tendon vibration remains effective under faster movements and with objects of different weights.

As a hint from the analysis of Experiment 2, Fig. 9 (Q3 and Q4 in (A)) indicates that resistance sensations are closely associated with the sense of heaviness, suggesting that perceived heaviness could be more reliably modulated by increasing or decreasing resistance during movement.

Furthermore, if the onset of the movement is crucial, stimulation should be applied ahead of the motion to account for the latency of the illusion.

### D. Toward integration in VR experiences

Our approach can be applied to virtual heaviness displays. Hirao and his colleagues have proposed combining pseudo-haptics with tendon vibration to simulate virtual heaviness [51]. They demonstrated that tendon vibration had a "noise" effect rather than a distinct effect in enhancing visual heaviness. On one hand, prior work combining tactile, visual, and proprioceptive stimuli reported that the illusion was enhanced [31]. On the other hand, previous studies have shown that interruptions in vision or other modalities modulated the illusion [68], [69]. Thus, congruency among sensory modalities might be necessary.

While Hirao et al. [51] extensively explored the effect of tendon vibration on pseudo-haptics, they consistently stimulated flexors or extensors during the lifting motion, the same as in this paper. If the congruency between visual and proprioceptive stimulation is crucial, other stimulation methods might induce an explicit effect. Furthermore, they used a lightweight controller (126 g) as a reference weight. The perception of heaviness through tactile sensations and proprioception can vary depending on the weight. van Beek et al. reported that "*Kinesthetic information was less reliable for lighter weights, while both sources of information were equally reliable for weights up to 300g*" [55]. Minamizawa et al. also found that tactile information is more effective for recognizing lighter objects, whereas kinesthetic information is more effective for recognizing heavier ones [70]. These findings suggest that original tactile information led to perceptual incongruencies, causing tendon vibration to act as "noise". Therefore, the influence of tactile information—through additional stimulation or different reference weights—should be further investigated.

## VIII. CONCLUSION

In this paper, we investigated whether tendon vibration on the ventral or dorsal side of the arm can increase or decrease the perceived heaviness while lifting. As a result of Experiment 1, a statistically significant enhancement and no significant reduction in perceived heaviness were observed, while some participants reported a decrease in the sense of heaviness. Focusing on the enhancement of the heaviness perception, Experiment 2 demonstrated that the increase in perceived heaviness can be at least adjusted in three levels in the range from 350 g to 450 g by varying the amplitudes.

While tendon vibration modulated the muscle sensations and changed the perceived heaviness, the mechanism underlying our approach remains to be determined. Primarily, the effect of vibrotactile stimulation needs further investigation. Also, the stimulation methods should be refined to improve the decrease effect in the future.

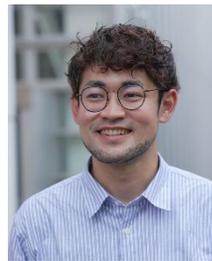

**Keigo Ushiyama** (Member, IEEE) received the Ph.D. degree in engineering from the University of Electro-Communications, Chofu, Japan. He is currently a project researcher at the University of Tokyo. His research interests include tactile displays, proprioceptive displays, and human interfaces.

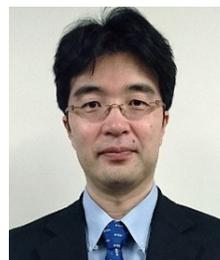

**Hiroyuki Kajimoto** (Member, IEEE) received the Ph.D. degree in information science and technology from The University of Tokyo, Tokyo, Japan, in 2004. He is currently a Professor with the Department of Informatics, University of Electro-Communications, Chofu, Japan. His research interests include tactile displays, tactile sensors, human interface, and virtual reality.